\documentclass[%
reprint,
superscriptaddress,
frontmatterverbose, 
showpacs,preprintnumbers,
longbibliography,
amsmath,amssymb,
aps,
pre,
]{revtex4-1}

\usepackage{graphicx}
\usepackage{dcolumn}
\usepackage{bm}
\usepackage{amsmath}

\renewcommand{\dim}{\mathcal{D}}
\newcommand{\pr}[1]{\left( #1 \right)}
\newcommand{\br}[1]{\left[ #1 \right]}

\newcommand{\braket}[1]{\langle #1 \rangle}

\begin{document}

\title{Geometric Evolution of Complex Networks}
\author{Charles Murphy}%
\email{charles.murphy.1@ulaval.ca}
\affiliation{D\'epartement de Physique, de G\'enie Physique, et d'Optique, Universit\'e Laval, Qu\'ebec (Qu\'ebec), Canada G1V 0A6}
\author{Antoine Allard}%
\affiliation{D\'epartement de Physique, de G\'enie Physique, et d'Optique, Universit\'e Laval, Qu\'ebec (Qu\'ebec), Canada G1V 0A6}
\affiliation{Centre de Recerca Matem\`atica, Edifici C, Campus Bellaterra, E-08193 Bellaterra (Barcelona), Spain}
\author{Edward Laurence}%
\affiliation{D\'epartement de Physique, de G\'enie Physique, et d'Optique, Universit\'e Laval, Qu\'ebec (Qu\'ebec), Canada G1V 0A6}
\author{Guillaume St-Onge}%
\affiliation{D\'epartement de Physique, de G\'enie Physique, et d'Optique, Universit\'e Laval, Qu\'ebec (Qu\'ebec), Canada G1V 0A6}
\author{Louis J. Dub\'e}%
\email{louis.dube@phy.ulaval.ca}
\affiliation{D\'epartement de Physique, de G\'enie Physique, et d'Optique, Universit\'e Laval, Qu\'ebec (Qu\'ebec), Canada G1V 0A6}
\date{\today}
\begin{abstract}
    We present a general class of geometric network growth mechanisms by homogeneous attachment in which the links created at a given time $t$ are distributed homogeneously between a new node and the exising nodes selected uniformly. This is achieved by creating links between nodes uniformly distributed in a homogeneous metric space according to a Fermi-Dirac connection probability with inverse temperature $\beta$ and general time-dependent chemical potential $\mu(t)$. The chemical potential limits the spatial extent of newly created links. Using a hidden variable framework, we obtain an analytical expression for the degree sequence and show that $\mu(t)$ can be fixed to yield any given degree distributions, including a scale-free degree distribution. Additionally, we find that depending on the order in which nodes appear in the network---its \textit{history}---the degree-degree correlation can be tuned to be assortative or disassortative. The effect of the geometry on the structure is investigated through the average clustering coefficient $\braket{c}$. In the thermodynamic limit, we identify a phase transition between a random regime where $\langle c \rangle \rightarrow 0$ when $\beta < \beta_\mathrm{c}$ and a geometric regime where $\langle c \rangle > 0$ when $\beta > \beta_\mathrm{c}$.
\end{abstract}
%
%
%
%
%
%
\maketitle
\section{Introduction} 
\label{sec:introduction}

Random geometric graphs (RGGs) provide a realistic approach to model real complex networks. In this class of models, $N$ nodes are located in a metric space $\mathcal{M}$ and are connected if they are separated by a distance shorter than a given threshold distance $\mu$ \cite{dall2002random,penrose2003random}. While RGGs are naturally associated with spatial networks \cite{barthelemy2011spatial}---such as infrastructure \cite{albert2004structural}, transport \cite{li2004statistical,guimera2004modeling,cardillo2006structural}, neuronal networks \cite{bullmore2012economy} and ad-hoc wireless networks \cite{waxman1988routing,kuhn2003ad,haenggi2009stochastic} ---, they can also be used to model real networks with no \textit{a priori} geographical space embedding. In these geometric representations, nodes are positioned in a hidden metric space where the distances between them encode their probability of being connected \cite{papadopoulos2011popularity,papadopoulos2015network1,papadopoulos2015network2}. This modeling approach allows to reproduce a wide range of topological properties observed in real networks, such as self-similarity \cite{serrano2008self}, high clustering coefficient \cite{krioukov2016clustering}, scale-free degree distribution \cite{krioukov2009curvature,krioukov2010hyperbolic}, efficient navigability \cite{boguna2009navigability} and distribution of weights of links \cite{allard2017geometric}.

This network geometry approach has been generalized to incorporate network growth mechanisms to further explain the observed structure of real networks under simple principles \cite{flaxman2006geometric,flaxman2007geometric,ferretti2011preferential,papadopoulos2011popularity,papadopoulos2015network1}. Two classes of mechanisms are considered in these approaches. The first one corresponds to a direct generalization of the classical preferential attachment (PA) coupled with a geometric mechanism: spatial or geometric preferential attachment \cite{flaxman2006geometric,flaxman2007geometric,ferretti2011preferential}. In this class of models, nodes are added on a manifold at each time $t$, similarly to a geometric prescription, but connect with the existing nodes with a probability proportional to their degree and to a distance dependent function $f(d)$. However, the power-law behavior of the degree distribution remains robust to the choice of a specific $f(d)$ and the curvature of space.

The second class involves the interplay between two attractiveness attributes, popularity and similarity, which dictates the connection probability. Contrary to spatial PA, new nodes connect more likely to high popularity, denoted by a hidden variable $r(t)$ dependent upon the time of birth $t$ of nodes, and to high similarity, denoted by the angular distance between two nodes positioned on a circle. This model is usually referred to as a spatial growing network model with \emph{soft} preferential attachment. Its evolution mechanism induces the power-law behavior of the degree distribution, but connects proportionally to their expected degree instead of their real degree. These network models have an interesting correspondence with static RGGs in hyperbolic geometry \cite{krioukov2009curvature,krioukov2010hyperbolic,papadopoulos2011popularity,ferretti2014duality}. This suggests that the hidden space of real networks might be hyperbolic as well.

While the PA mechanism and hyperbolic geometry have been proved to naturally generate networks with power-law degree distribution, they do not capture the whole range of fundamental structural properties characterizing real networks. A good example is the assortative behavior of certain social networks such as scientific collaborations networks \cite{newman2001scientific,newman2002assortative,newman2003mixing1}, film actor networks \cite{amaral2000classes} and Pretty good privacy (PGP) web of trust networks \cite{boguna2004models}. The reason for the lack of assortativity in the PA and hyperbolic models is that they map to the \emph{soft configuration model}: a maximum entropy ensemble in which the degree sequence is fixed with soft constraints such that no degree-degree correlation can be enforced \cite{krioukov2010hyperbolic,zuev2016hamiltonian}. Growth mechanisms with custom degree-degree correlations are therefore still wanting.

We present a type of growing geometric network which, in contrast with PA, distributes the links created by newborn nodes \emph{homogeneously} among the existing ones. We call this attachment process \emph{homogeneous attachment} (HA). From a geometric point of view, HA is interpreted as a growing geometric network mechanism where the connection threshold $\mu \equiv \mu(t)$ is a general function of the time of birth of the newborn node. This feature allows the creation of an arbitrary number of links at each time enabling direct specification of the degree distribution \emph{and} the degree-degree correlations.

The paper is organized as follows. In Sec.~\ref{sec:growing_geometric_networks}, the growing geometric network model is presented in detail. Section~\ref{sec:degree_sequence} is devoted to the development of an analytical expression for the degree of each node. This analytical description fixes $\mu(t)$ for any type of degree sequences, and therefore specifies the degree distribution. In Sec.~\ref{sec:network_history}, we show how the history of a network (the order of appearance of nodes) can be used to tune the degree-degree correlations without altering the degree distribution. In Sec.~\ref{sec:geometry_effetcs}, the effects of the underlying geometry on the network topology are studied with special attention given to the average clustering coefficient $\braket{c}$. Finally, in Sec.~\ref{sec:conclusion} we draw some conclusions, limitations of the model and perspectives.

%
\section{Growing geometric networks} 
\label{sec:growing_geometric_networks}

\begin{figure}
    \centering
    \includegraphics[scale=0.8]{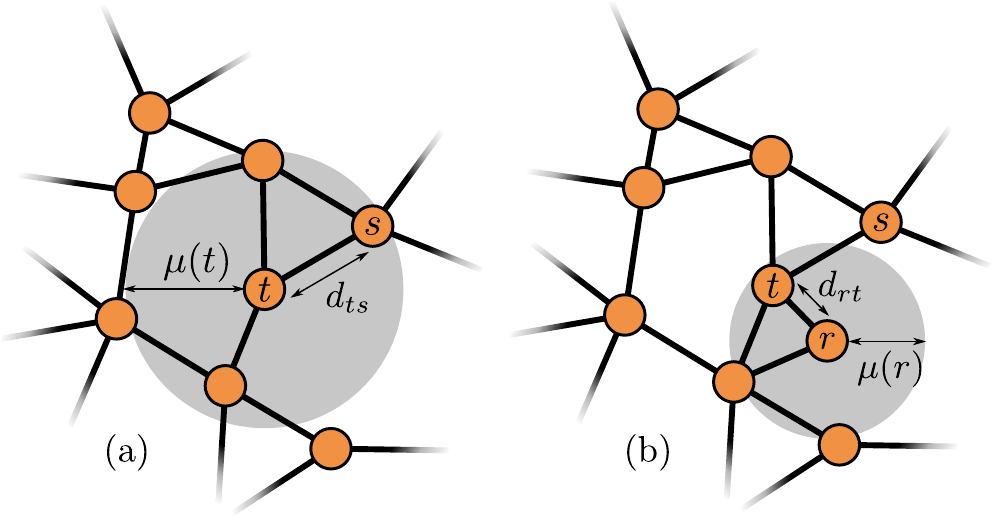}
    \caption{Illustration of the HA growth mechanism for geometric networks: (a) birth of nodes $t$ which connects with three neighbors, including $s$ at distance $d_{ts} < \mu(t)$, (b) subsequent birth of node $r>t$ connecting to $t$  with $d_{rt} < \mu(r)$. For illustration purposes, $p[x_t, x_s; \mu(t)] = \Theta[\mu(t) - d_{ts}]$ such that any node within a circle of radius $\mu(t)$ (gray area) centered on node $t$ will become connected to it..}
    \label{fig:GRN_scheme}
\end{figure}

Let us consider the isotropic, homogeneous and borderless surface of a $(\mathcal{D}+1)$-ball of radius $R$ as the metric space $\mathcal{M}$ (dimension $\dim$) in which the growing geometric networks are embedded. This choice simplifies the analytical calculations below, but does not alter the generality of our conclusions. Considering an initially empty metric space, the growing process goes as follows (see Fig.~\ref{fig:GRN_scheme}):
\begin{enumerate}
    \item At any time $t\geq1$, a new node (noted $t$) is assigned the random position $x_t$ uniformly distributed on $\mathcal{M}$.
    \item Node $t$ connects with the existing nodes $s<t$ with probability $p(x_t, x_s; t)$.
    \item Steps 2 and 3 are repeated until a total of $N$ nodes have been reached.
\end{enumerate}

In the model, $p(x, y; t)$ is a general function of the birth time $t$ and the positions of the nodes $x$ and $y$ that we leave unspecified for the moment. Notice, however, that both the spatial and time dependencies have nontrivial effects on the network topology. On the one hand, the geometry encoded in $p(x, y; t)$ will affect the properties of the networks like the distribution of component sizes \cite{dettmann2016random} and the clustering coefficient \cite{krioukov2016clustering}. On the other hand, the time dependency will determine \textit{when} distant connections are allowed which, in turn, induce correlations between nodes with different birth times. For instance, if $p(x, y; t)$ is a decreasing function of $t$, older nodes will on average be hubs tighly connected to one another while younger nodes will have lower degrees. The choice of $p(x, y; t)$ can therefore induce a hierarchical structure typical of assortative networks, where the hubs are at the topological center of the structure and the low degree nodes occupy its outskirt.

A natural and straightforward generalization of previous works \cite{dall2002random,penrose2003random} is to add a time dependence to the probability of connection
\begin{equation}\label{eq:heaviside_prob}
    p\big[x, y; \mu(t)\big] = \Theta[\mu(t) - d(x, y)] = \begin{cases} 1 & \text{if } d(x, y) < \mu(t) \\ 0 & \text{otherwise} \end{cases}\,,
\end{equation}
where $\Theta(x)$ is the Heaviside step function, $d(x, y)$ is the metric distance between $x$ and $y$ and $\mu(t)$ is the connection threshold. Fixing $\mu(t) = \mu$ reduces to the known \textit{sharp} RGG model which is deterministic in the creation of the links but not always suitable to describe real networks \cite{waxman1988routing,kuhn2003ad,balister2004continuum}. 

For more flexibility, we consider a connection probability analog to the Fermi-Dirac distribution
\begin{equation}\label{eq:fermi_dirac_prob}
    p\big[x, y; \mu(t), \beta\big] = \frac{1}{\exp\left\{\beta\big[d(x, y) - \mu(t)\big]\right\} + 1}\,,
\end{equation}
where $\beta$ is a parameter controlling the clustering coefficient and $\mu(t)$ limits the spatial extent of new links \cite{krioukov2009curvature,krioukov2016clustering}. From a statistical physics point of view, using this connection probability amounts to consider the links as fermions of energy given by the length of the links, $d(x, y)$, embedded in an environment maintained at temperature $1/\beta$ with a chemical potential $\mu(t)$.

This connection probability is interesting for two reasons. First, it is very similar to the probability of connection of the exponential random graph model. The ensuing network ensemble maximizes the Gibbs entropy when the average number of links between any given pair of nodes is fixed \cite{park2004statistical}. Second, varying $\beta$ enables us to navigate between the hot regime $\beta\to0$ and the cold regime $\beta\to \infty$. In the limit $N \rightarrow \infty$ and when $\mu(t) = \mu$, the connection probability in the hot regime no longer depends on the position of the nodes, and the corresponding network ensemble is of the Erd\H{o}s-R\'enyi type, where $\langle c \rangle = \mathcal{O}(N^{-1}) \rightarrow 0$ \cite{krioukov2016clustering}. In contrast, in the cold regime, under the same conditions, $\braket{c}$ reaches a maximum independent of $N$ \cite{krioukov2016clustering}.

%
\section{Degree Sequence} 
\label{sec:degree_sequence}

Since $\mu(t)$ limits the spatial extent of potential connections, it has a direct impact on the degree sequence. In this section, we shed light on the relation between $\mu(t)$ and the resulting structure.

\subsection{Hidden variables} 
\label{sub:hidden_variables}

A convenient way to analyze the HA mechanism is via the framework of random graphs with hidden variables \cite{boguna2003class}. In this ensemble, each node $i = 1, \cdots, N$ is assigned a hidden variable $h_i$, sampled from a probability distribution $\rho(h)$, and links are created between nodes $(i, j)$ with probability $p_H(h_i, h_j)$. This general model encodes the correlation among nodes via the hidden variables, which can either be random numbers $h_i$ or vectors of random numbers $\bm{h}_i$. Although this model is very general and versatile, it is nevertheless amenable to a full mathematical description of the structural properties of the network ensemble such as the degree distribution, the correlations and the clustering.

In our model, there are two hidden variables involved: the time of birth $t$ and the position $x$ on $\mathcal{M}$. Whereas $x$ is a random variable distributed uniformly on $\mathcal{M}$ by definition, the same cannot be said straightforwardly for $t$. However, since a node was born at every $t=1, \ldots, N$, randomly choosing a node of birth time $t$ will occur uniformly. This is sufficient to provide a mathematical description of the network ensemble using the hidden variable framework. The hidden variable probability distribution is then 
\begin{equation}
    \rho(\bm{h}) = \frac{1}{N S_\dim(R)} = \frac{\Gamma\left(\frac{\dim+1}{2}\right)}{N 2 \pi^{(\dim+1)/2} R^{\dim}} 
\end{equation}
where $S_\dim(R)$ is the surface of the $(\dim + 1)$-ball of radius $R$ and $\Gamma(x) = \int_0^\infty t^{x - 1} e^{-t} dt$ is the Gamma function. 

For the sake of simplicity, we consider the characteristic time $T \equiv N$ and length $X \equiv \pi R$ to define the normalized hidden variable $\bm{\tilde{h}} = (\tau, \xi)$, where $\tau \equiv \tau(t) = \frac{t}{T} = \frac{t}{N}$, $\xi \equiv \xi(x) = \frac{x}{X} = \frac{x}{\pi R}$, and the normalized quantities $\tilde{R} \equiv \frac{R}{X} = \frac{1}{\pi}$, $\tilde{\mu}(\tau) \equiv\frac{\mu(\tau T)}{X} = \frac{\mu (\tau N)}{\pi R}$ and $\tilde{\beta} \equiv\beta X = \beta \pi R$. This choice of normalization implies that $\rho(\bm{\tilde{h}}) = \frac{\Gamma\left(\frac{\dim + 1}{2}\right)}{2 \pi^{(1- \dim)/2}} \equiv \rho$ becomes a constant. Note also that as $N$ tends to infinity, the difference between birth times, $\Delta t = N^{-1}$, tends to zero which allows to consider the birth times $\tau$ as a continuous variable in $(0, 1]$ to facilitate the analytical calculations. Additionally, because $\mathcal{M}$ is homogeneous and isotropic, all analytical calculations will always consider a node positioned at the origin $\bm{0}$ without loss of generality. Finally, considering two nodes with hidden variables $\bm{h} = (\xi, \tau)$ and $\bm{h'} = (\xi', \tau')$, the connection probability becomes

\begin{equation}
    \begin{split}
        p_H(\bm{\tilde{h}}, \bm{\tilde{h}}') &= \Theta(\tau - \tau') p\big[\xi, \xi'; \tilde{\mu}(\tau), \tilde{\beta}\big] \\ &+ \Theta(\tau' - \tau) p\big[\xi', \xi; \tilde{\mu}(\tau'), \tilde{\beta}\big]\,.
    \end{split}
\end{equation}
where the step functions select the appropriate probability depending on which nodes appeared first.


\subsection{Fixing the degree sequence} 
\label{sub:fixing_the_degree_sequence}

Since the new links are distributed homogeneously among the existing nodes at any given time, the expected degree of node $\tau$ \emph{at the end of the process} has the following simple form
\begin{equation}\label{eq:degree_sequence}
    \kappa(\tau) = N \bigg[\tau n(\tau) + \int_\tau^1 n(\sigma) d \sigma\bigg]\,,
\end{equation}
where
\begin{equation}\label{eq:fraction_of_created_links}
    n(\tau) = \int_\mathcal{M}  p[\bm{0}, \xi; \tilde{\mu}(\tau), \beta] \rho \, d \xi
\end{equation}
which corresponds to the probability that node $\tau$ will connect to any existing nodes. This integral can be solved analytically when $\dim = 1$ leading to a closed form expression,
\begin{equation}
    n(\tau) = \frac{1}{\tilde{\beta}} \ln \left\{\frac{\exp\pr{-\tilde{\beta}} + \exp\br{-\tilde{\beta}\tilde{\mu}(\tau)}}{1 + \exp\br{-\tilde{\beta}\tilde{\mu}(\tau)}}\right\} \,.
\end{equation}
For $\dim \neq 1$, Eq.~\eqref{eq:fraction_of_created_links} must be solved numerically. Each term of Eq.~\eqref{eq:degree_sequence} can be interpreted explicitly: the first one corresponds to the number of links which node $\tau$ creates on its arrival while the second term accounts for the links it gains by the creation of the other nodes.

The average degree of node $\tau$ can be obtained as a function of $\tilde{\mu}(\tau)$ using Eq.~\eqref{eq:fraction_of_created_links}. Indeed, Eq.~\eqref{eq:degree_sequence} can be inverted such that we obtain $n(\tau)$ as a function of the degree sequence $\left\{\kappa(\tau)\right\}$. We start by deriving Eq.~\eqref{eq:degree_sequence} with respect to $\tau$ such that we obtain the following differential equation
\begin{equation}
    \frac{d \kappa(\tau)}{d \tau} = N \tau \frac{d n(\tau)}{d \tau}\,.
\end{equation}
Then, solving for $n(\tau)$ by integrating by part from $\tau$ to 1, this yields
\begin{equation}\label{eq:inversion_k_to_n}
    n(\tau) = \frac{1}{N}\br{\frac{\kappa(\tau)}{\tau} - \int_\tau^1 \frac{\kappa(\sigma)}{\sigma^2}d \sigma }\,. 
\end{equation}
This represents one of the most interesting assets of the HA model: given an ordered degree sequence in time $\kappa(\tau)$, it is possible to calculate analytically $n(\tau)$ to obtain the appropriate form of $\tilde{\mu}(\tau)$ via Eq.~\eqref{eq:fraction_of_created_links} which, in the case $\dim = 1$, yields
\begin{equation}\label{eq:scalefree_chemicalpotential}
    \tilde{\mu}(\tau) = \frac{1}{\tilde{\beta}} \ln \left\{\frac{1 - \exp\br{-\tilde{\beta} n(\tau)}}{\exp\br{-\tilde{\beta} n(\tau)} - \exp\pr{-\tilde{\beta}}}\right\} \, .
\end{equation}
It is therefore possible to obtain the corresponding function $\tilde{\mu}(\tau)$ to reproduce any given degree sequence.

It is important to understand at this point that what is fixed here is the ordered expected degree sequence in time, which is quite different from a configuration model prescription. For example, an increasing degree sequence in time would yield a totally different structure than a decreasing one. We will show that this additional trait of the model allows to tune the level of correlations between the degree of nodes.


\subsection{Scale-Free Growing Geometric Networks} 
\label{sub:scale_free_growing_geometric_networks}

Scale-free networks ($P(k) \propto k^{-\gamma}$) can be generated with our model by assuming that the ordered degree sequence has the following form
\begin{equation}\label{eq:scalefree_degreesequence}
    \kappa(\tau) = \nu \tau^{-\alpha}\,,
\end{equation}
where $\nu > 0$ fixes the average degree $\braket{k}$, and where $0<\alpha \leq 1$ controls the exponent of the degree distribution via $\gamma = 1 + \frac{1}{\alpha}$. It is then possible to calculate explicitly $n(\tau)$ using Eq.~\eqref{eq:inversion_k_to_n}
\begin{equation}\label{eq:scalefree_fraction}
    n(\tau) = \frac{\nu}{N(\alpha + 1)} \pr{\alpha \tau^{-\alpha -1} + 1}\,.
\end{equation}

\begin{figure}
    \centering
    \includegraphics[scale=0.6]{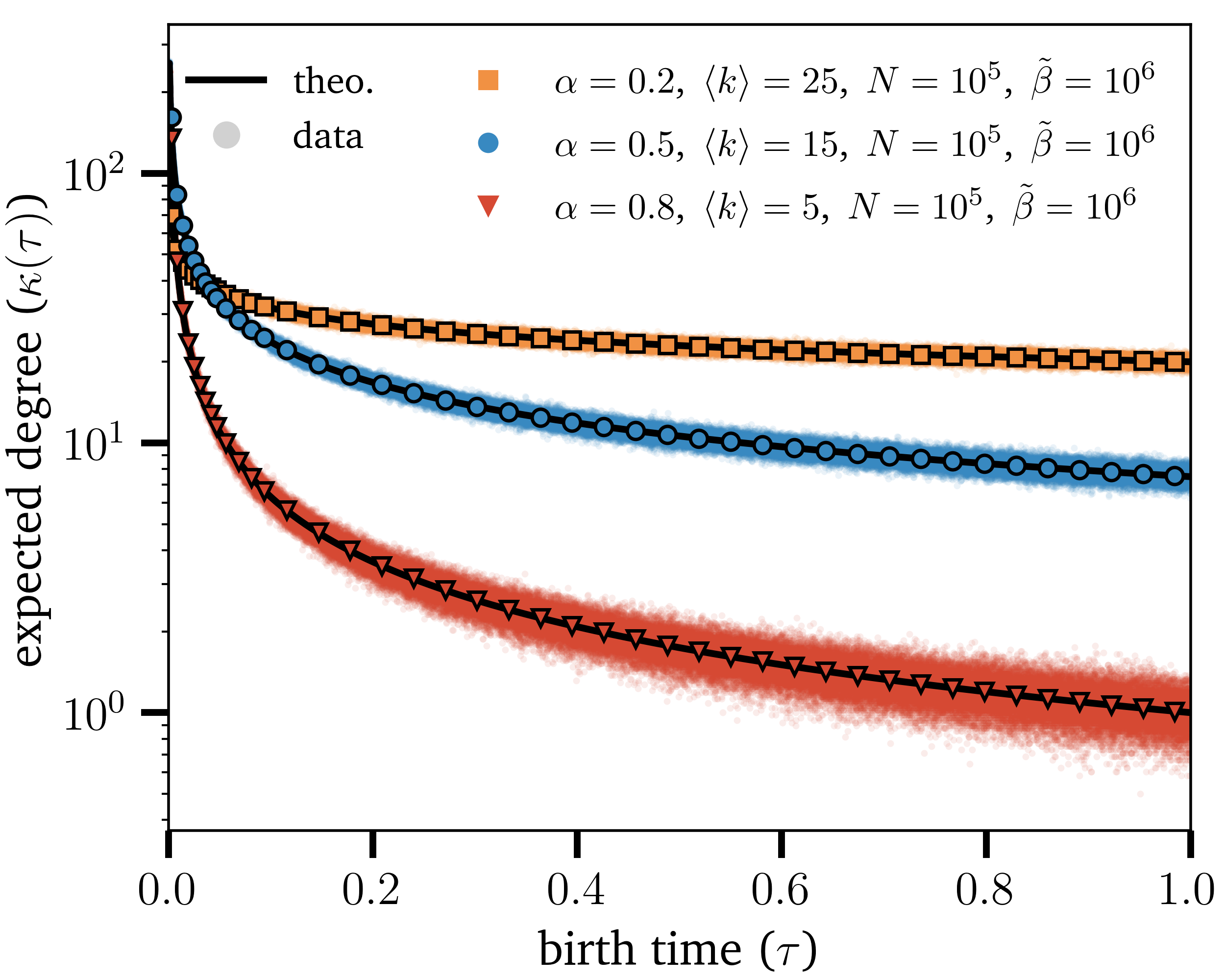}
    \caption{(Color online) $\kappa(\tau)$ as a function of the birth time $\tau$ for different parameters. The small dots correspond to data obtained from Monte Carlo simulations averaged over 100 instances where $\tilde{\mu}(\tau)$ is given by Eq.~\eqref{eq:scalefree_chemicalpotential} with $n(\tau)$ given by Eq.~\eqref{eq:scalefree_fraction}. The markers (squares, circles and triangles) correspond to the average data. The black solid lines corresponds to Eq.~\eqref{eq:scalefree_degreesequence}. The parameters of each dataset are indicated on the plot. The networks have been averaged over 96 instances.}
    \label{fig:k_vs_tau}
\end{figure}
\begin{figure}
    \centering
    \includegraphics[scale=0.6]{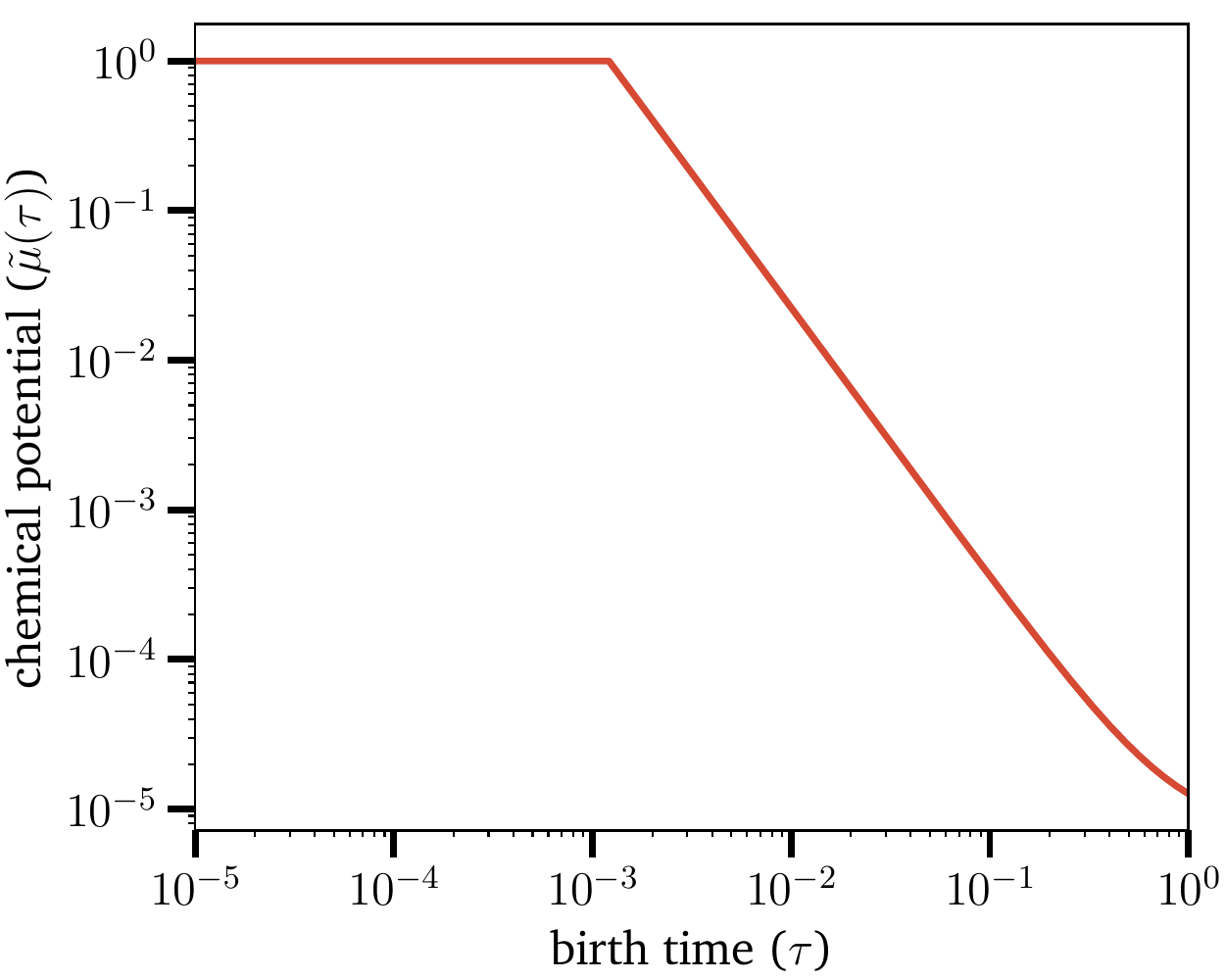}
    \caption{Chemical potential $\tilde{\mu}(\tau)$ vs $\tau$ required to generate scale-free geometric networks with $\alpha = 0.8$, $\braket{k}=5$, $N=10^5$ and $\tilde{\beta}=10^7$. The line corresponds to Eq.~\eqref{eq:scalefree_chemicalpotential} with $n(\tau)$ given by Eq.~\eqref{eq:scalefree_fraction}.}
    \label{fig:mu_vs_tau}
\end{figure}

Figure~\ref{fig:k_vs_tau} shows $\kappa(\tau)$ for scale-free geometric networks with different sets of parameters where the chemical potential is given by Eq.~\eqref{eq:scalefree_chemicalpotential} (see Fig.~\ref{fig:mu_vs_tau}). Clearly, Eq.~\eqref{eq:scalefree_degreesequence} correctly describes the behavior of the degree sequence.


\subsection{Finite-size effects} 
\label{sub:finite_size_effects}

The agreement between the theoretical predictions and the numerical simulations demonstrates it is indeed possible to reproduce any degree distributions with an appropriate choice of $\tilde\mu(\tau)$. There are some limitations however. It is possible for the ordered sequence of expected degrees to be such that Eq. (9) yields $n(\tau) > 1$ meaning that node $\tau$ whould have to create more links than the number of already existing nodes at the moment of its birth (time $\tau$). In the case of scale-free networks with an ordered degree sequence of the form of Eq.~\eqref{eq:scalefree_degreesequence}, this happens for all nodes $\tau < \tau^*$ where
\begin{equation}\label{eq:old_core}
    \tau^* = \br{\frac{\alpha \nu}{N(\alpha + 1) - \nu}}^{\frac{1}{\alpha + 1}}\,,
\end{equation}
which corresponds to the plateau seen on Fig.~\ref{fig:mu_vs_tau}, and implies that all nodes $\tau < \tau^*$ will form a connected clique of hubs. Limiting $n(\tau)$ such that $n(\tau) < 1$ further implies that some links will be missing, and consequently that the degree sequence will differ from the one given by Eq.~\eqref{eq:scalefree_degreesequence}. This discrepancy can be investigated through the average degree which, without this correction, is equal to $\langle k \rangle = \frac{\nu}{1 - \alpha}$. Considering the clique of connected hubs yields instead
\begin{align}
    \braket{k} &= \int_0^1 \kappa(\tau) d \tau \notag \\
    &= \tau^* \kappa(\tau^*) + \int_{\tau^*}^1 \kappa(\tau) d \tau \notag\\
    &= \frac{\nu}{1 - \alpha} \pr{1 - \alpha {\tau^*}^{1 - \alpha}}\,.
\end{align}
Thus, the difference between the two values for the average degree scales as $\mathcal{O}(N^{-\frac{1-\alpha}{1+\alpha}})$ which vanishes in the limit $N \rightarrow \infty$. One possible way to circumvent this effect would be to allow multilinks and self-loops, but this is left as a future improvement of the model.


%
\section{Network History} 
\label{sec:network_history}

The degree sequence can take many forms: choosing Eq.~\eqref{eq:scalefree_degreesequence} for the degree sequence yields scale-free networks. Yet, this is only one example of degree sequence capable of generating networks with a power-law degree distribution. Another example of such degree sequence would be to choose all the entries of $\kappa(\tau)$ at random from a distribution $P(k) \propto k^{-\gamma}$. However, these two growth processes, despite having the same degree distribution, have different $\tilde{\mu}(\tau)$ which in turn affect the structural organization of the generated networks.

Reordering the degree sequence amounts to changing the \emph{history} of the network. We define a history by a set $H = \left\{\tau_i\right\}$ of the birth time, where $\tau_i$ is the birth time of a node with label $i$ of fixed final degree $k_i$. For a specific history, the final expected degree of node $i$ is simply
\begin{equation}
    \kappa(\tau_i) = k_i \,.
\end{equation}
A network where the degree sequence $\left\{k_i\right\}$ is known can have different histories. Unlike most network growth models, in HA the degree sequence is preserved even if the network history is changed. This particularity makes HA a unique alternative to model growing networks because it induces a specific correlation between nodes born at a different times. In other words, the change of network history affects the correlation between nodes. 

\subsection{Degree-Degree Correlation}

To quantify the impact of the history on the resulting network structure, we consider the degree-degree correlation. This measure is fully characterized by the conditional probability that a node of degree $k$ is connected to another node of degree $k'$ denoted by $P(k'|k)$.

We express the degree-degree correlation in terms of the birth times as the conditional probability $p(\sigma|\tau)$ that node $\tau$ is connected to node $\sigma$ given by
\begin{equation}
    p(\sigma|\tau) = \frac{N}{\kappa(\tau)}\bigg[\Theta(\tau - \sigma)n(\tau)  + \Theta(\sigma - \tau) n(\sigma)\bigg]\,.
\end{equation}
However, it is usually more convenient to calculate the average degree of nearest-neighbors (ANND) denoted by $\kappa_{nn}(\tau)$ and defined by
\begin{align}\label{eq:ANND_time}
    \kappa_{nn}(\tau) &= \int_{0}^1 \kappa(\sigma) p(\sigma | \tau) d \sigma \notag\\
    &= N\br{\int_{0}^\tau \frac{\kappa(\sigma) n(\tau)}{\kappa(\tau)} d \sigma + \int_{\tau}^1 \frac{\kappa(\sigma) n(\sigma)}{\kappa(\tau)} d \sigma}\,.
\end{align}
From this expression, the degree-dependent ANND, denoted $\bar{k}_{nn}(k)$, can be obtained via the hidden variable framework
\begin{equation}\label{eq:ANND_degree}
    \bar{k}_{nn}(k) = 1 + \frac{1}{P(k)} \int_0^1 g(k|\tau) \kappa_{nn}(\tau) d \tau \,,
\end{equation}
where $g(k|\tau) = \frac{e^{-\kappa(\tau)} \kappa(\tau)^{k}}{k!}$ \cite{boguna2003class}. Having this analytical expression in hand, we can now investigate the effect of different histories on the degree-degree correlations.

\begin{figure*}
    \centering
    \includegraphics[scale=0.6]{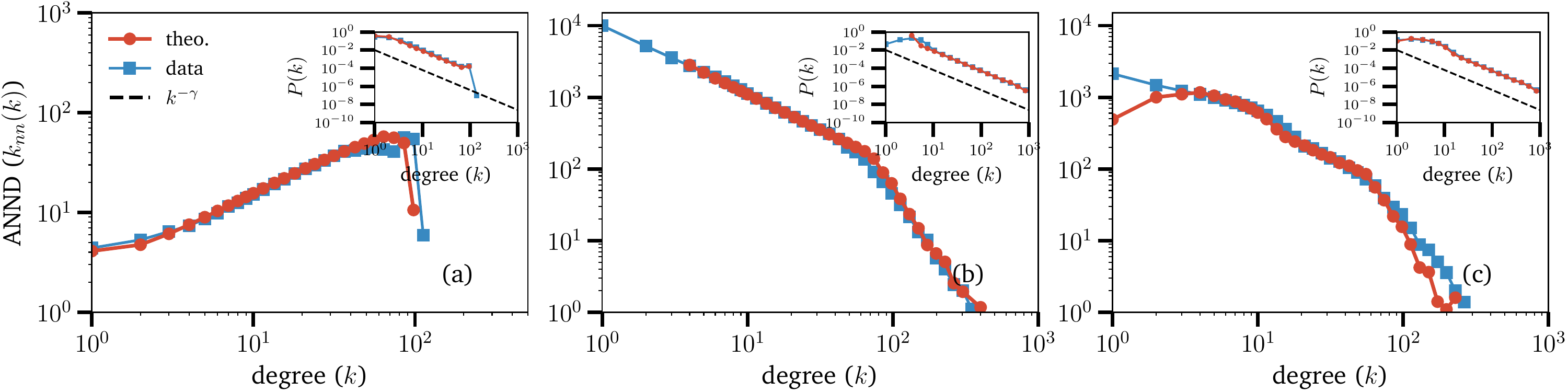}
    \caption{(Color online) Degree-degree correlation for different histories: (a) decreasing degree order (Sec.~\ref{sub:decreasing_degree_order}), (b) increasing degree order (Sec.~\ref{sub:increasing_degree_order}), (c) random order (Sec.~\ref{sub:random_order}). In each plot, the results from Monte Carlo simulations are indicated by the blue squares and the analytics, by the red circles. The degree distributions are also displayed in inset and are power laws with $\gamma=2.2$ (dashed lines). The average degree assortativity coefficient $\braket{r}$ amounts to (a) $\braket{r} \simeq 0.70$, for (b) $\braket{r} \simeq -0.19$ and for (c) $\braket{r} \simeq -0.14$. For each simulation, $\alpha = 0.83$, $\braket{k} = 6$, $N = 10^4$, $\tilde{\beta} = 10^6$. The networks have been averaged over 48 instances.}
    \label{fig:knn_vs_k_orders}
\end{figure*}

\subsection{Decreasing Degree Order}
\label{sub:decreasing_degree_order}

As we have seen in Sec.~\ref{sec:degree_sequence}, scale-free degree sequences can be written as $\kappa(\tau) = \nu \tau^{-\alpha}$. This implies a specific type of network history where the degree sequence is a decreasing order of the degrees: hubs are old while low degree nodes are young. The birth time correlations can then be calculated using Eq.~\eqref{eq:scalefree_degreesequence} and \eqref{eq:scalefree_fraction}
\begin{equation}
    \kappa_{nn}(\tau) = \frac{\kappa(\tau)}{2 (1 - \alpha)}\left[1 + \tau^{2\alpha}\right] \,,
\end{equation}
which is well approximated, for $\tau\ll1$, by $\kappa_{nn}(\tau) \simeq \frac{\kappa(\tau)}{2(1 - \alpha)}$. Then, the $\bar{k}_{nn}(k)$ can be calculated and is given, for large $k$, by
\begin{equation}\label{eq:ANND_scalefree_approx}
    \bar{k}_{nn}(k) \simeq \frac{k}{2(1 - \alpha)} \,.
\end{equation}
Since $\bar{k}_{nn}(k)$ is essentially a linear function of $k$, it shows that choosing an ordered by degree history yields assortative networks. Figure~\ref{fig:knn_vs_k_orders}(a) confirms these predictions.

\subsection{Increasing Degree Order}
\label{sub:increasing_degree_order}

We now consider an ordering in which the old nodes are assumed to be the low degree ones and the young nodes are the hubs. To generate scale-free networks with an increasing degree order, we use
\begin{equation}
    \kappa(\tau) = \nu (1 - \tau)^{-\alpha} \,,
\end{equation}
which yields
\begin{equation}
    n(\tau) = \frac{\nu}{N} \br{ \frac{(1 - \tau)^{-\alpha}}{\tau} - \int_\tau^{1} \frac{(1 - \sigma)^{-\alpha}}{\sigma^2} d \sigma} \,.
\end{equation}
As illustrated on Fig.~\ref{fig:knn_vs_k_orders}(b), an increasing degree history implies a decreasing $\bar{k}_{nn}(k)$ which corresponds to disassortative networks. This can be explained by the fact that, since all the low degree nodes are created early in the history, when the time comes for the hubs to be born, they will connect more frequently to them. Combined with the results of Section~\ref{sub:decreasing_degree_order}, these results show how the degree-degree correlation can be tuned simply by changing the history we consider.

\subsection{Random Order}
\label{sub:random_order}

Our last example is when the set $H$ is random. The network history is then composed of birth times $\tau_i$ that are random variables distributed uniformly between $[0,1)$. From the point of view of the degree sequence, this means that $\kappa(\tau)$, and consequently any function of $\tau$, is also a random variable but drawn, in the case of $\kappa(\tau)$, from the degree distribution. Because of this, $\kappa(\tau)$ is no longer a continuous function and Eqs.~\eqref{eq:inversion_k_to_n} and \eqref{eq:ANND_time} can no longer be used. Instead, we use the following discrete form
\begin{equation}\label{eq:fraction_discrete}
    n(\tau) = \frac{1}{N}\br{\frac{\kappa(\tau)}{\tau} - \sum_{i=t+1}^N \frac{\kappa(\tau_i)}{\tau_i^2} \Delta \tau_i} \,,
\end{equation}
where $t \equiv \tau N$ is the number of nodes when $\tau$ is born, $\tau_i$ is the birth time of the $i^{th}$ node to arrive in the network and $\Delta \tau_i = \tau_{i} - \tau_{i-1}$, with $\tau_{i} =0$ if $i\leq0$, is the time step between two birth events. In a similar way, the ANND can be adapted as well
\begin{equation}\label{eq:ANND_discrete}
    \kappa_{nn}(\tau) = N \br{\sum_{i=1}^t \frac{\kappa(\tau_i) n(\tau)}{\kappa(\tau)}\Delta \tau_{i} + \sum_{j=t+1}^N \frac{\kappa(\tau_j) n(\tau_j)}{\kappa(\tau)}\Delta \tau_{j}}\,.
\end{equation}

These expressions can be obtained by evaluating the integrals in Eq.~\eqref{eq:inversion_k_to_n} and \eqref{eq:ANND_time} in the form of Riemann sums. Therefore, in the thermodynamic limit, $\Delta \tau_i \to 0$ for all $i$ and the continuous and discrete forms are totally equivalent.

As before, to generate scale-free networks under this process, we would have to first determine the degree sequence $\kappa(\tau)\sim P(k) \propto k^{-\gamma}$ and then determine the corresponding $\tilde{\mu}(\tau)$. With that procedure one generates networks with a random history. 

As we can see on Fig.~\ref{fig:knn_vs_k_orders}(c), similarly to the increasing degree history, $\bar{k}_{nn}(k)$ is a decreasing function and the networks show disassortativity with $\braket{r} = -0.14$. However, the disassortativity observed here is entirely due to structural constraints imposed by the degree sequence: the degree sequence forces the hubs to connect more frequently to the low degree nodes.

%
\section{Geometry Effects} 
\label{sec:geometry_effetcs}

The conclusions drawn so far are general, whether the networks are geometric or not. The effect of the geometry becomes manifest at the level of the three-node correlations where the triangle inequality of the underlying metric space implies a non-vanishing clustering coefficient $\braket{c}$ in the thermodynamic limit. The choice of $p\big[\xi, \psi; \tilde{\mu}(\tau), \tilde{\beta}\big]$ as a Fermi-Dirac distribution [Eq.~\eqref{eq:fermi_dirac_prob}] allows us to adjust the level of clustering by changing the inverse temperature $\tilde{\beta}$.

The average clustering coefficient $C$ is the fraction of triplets ---three nodes connected in chains--- actually forming a triangle. Adapting this coefficient to each node instead of the whole network yields the local clustering coefficient $c$, where $c$ corresponds to the fraction of a node's neighbors that are connected. For node $\tau$, this fraction yields
\begin{equation}\label{eq:birthtime_clustering}
\begin{split}
    c(\tau) = \frac{2 N^2 \rho^2}{\kappa^2(\tau)}&\left[\int_0^\tau \sigma \ell(\tau, \sigma,\tau) d \sigma  + \tau\int_\tau^1 \ell(\tau,\sigma,\sigma) d \sigma \right. \\
   &+ \left. \int_\tau^1 \int_\tau^{\sigma} \ell(\sigma,\sigma, \rho) d \sigma d \lambda \right] \,,
\end{split}
\end{equation}
where $\ell(\tau,\sigma,\lambda)$ is the probability that nodes $\tau$, $\sigma$ and $\lambda$ form a triangle and is given by
\begin{equation}\label{eq:triangle_prob}
\begin{split}
    \ell(\tau,\sigma,\lambda) = \int_\mathcal{M} \int_\mathcal{M} &p[\bm{0}, \xi; \tilde{\mu}(\tau), \tilde{\beta}] p[\xi, \psi; \tilde{\mu}(\sigma), \tilde{\beta}] \\
    &\times p[\psi, \bm{0}; \tilde{\mu}(\lambda), \tilde{\beta}] \, d \xi d \psi \,.    
\end{split}
\end{equation}
Unfortunately, Eq.~\eqref{eq:birthtime_clustering} cannot be solved analytically for any $\tilde{\beta}$ and $\dim$. However, the limiting cases consisting of the cold ($\tilde{\beta}\to\infty$) and hot ($\tilde{\beta}\to0$) limit with $\dim = 1$ have closed forms for $c(\tau)$.

\subsection{Cold Limit $\tilde{\beta}\to\infty$}

\begin{figure}
    \centering
    \includegraphics{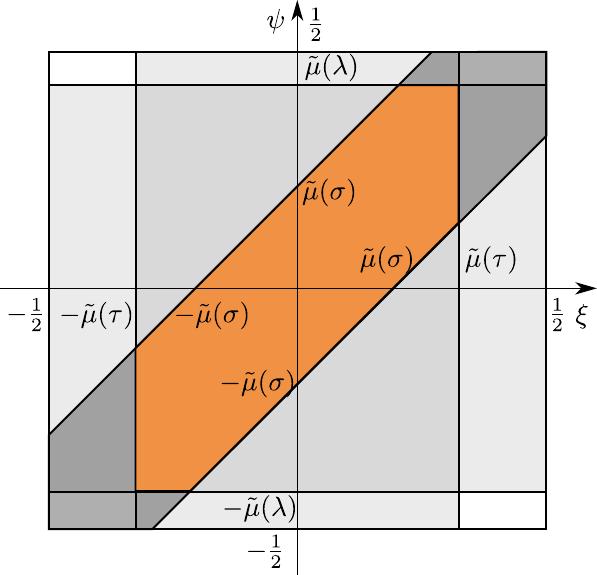}
    \caption{(Color online) Geometrical representation of Eq.~\eqref{eq:triangle_prob} in the limit $\tilde{\beta}\to\infty$. The constraints of the domain are illustrated in gray and the area in orange corresponds to the resulting value of Eq.~\eqref{eq:triangle_prob}. In this illustrative case, $\tilde{\mu}(\tau) > \tilde{\mu}(\sigma)$ and $\tilde{\mu}(\rho) > \tilde{\mu}(\sigma)$.}
    \label{fig:clustering_area}
\end{figure}

In this regime, the connection probability $p\big[\xi, \psi; \tilde{\mu}(\tau), \tilde{\beta}\big]$ takes the form of a Heaviside step function centered at $\tilde{\mu}(\tau)$, which maximizes the clustering coefficient, $\braket{c}$, when $\tilde{\mu}(\tau)$ does not depend on $\tau$ \cite{krioukov2016clustering}. Equation~\eqref{eq:scalefree_chemicalpotential} then yields
\begin{equation}\label{eq:cold_chemicalpotential}
    \tilde{\mu}(\tau) \simeq n(\tau) \,.
\end{equation}
Additionally, with $\dim = 1$, $\ell(\tau, \sigma, \lambda)$ can be calculated geometrically from the area of a truncated parallelogram (see Fig.~\ref{fig:clustering_area}). It yields
\begin{equation}\label{eq:cold_triangle}
\ell(\tau,\sigma,\lambda) \simeq
    \begin{cases}
        \ell^*(\tau,\sigma,\lambda), & \tilde{\mu} (\sigma) < \tilde{\mu}(\tau) + \tilde{\mu}(\lambda)\\
        4 \tilde{\mu}(\tau) \tilde{\mu}(\lambda), & \text{otherwise}.
    \end{cases}
\end{equation}
where
\begin{equation}
    \begin{split}
    \ell^*(\tau,\sigma,\lambda) \simeq & 2\br{\tilde{\mu}(\tau)\tilde{\mu}(\sigma) + \tilde{\mu}(\sigma)\tilde{\mu}(\lambda) + \tilde{\mu}(\lambda)\tilde{\mu}(\tau)} \\ &- \br{\tilde{\mu}^2(\tau) + \tilde{\mu}^2(\sigma) + \tilde{\mu}^2(\lambda)} 
    \end{split}
\end{equation}
From Eq.~\eqref{eq:cold_chemicalpotential}, we know that $\tilde{\mu}(\tau) = \mathcal{O}(N^{-1})$, which implies that $\ell(\tau,\sigma,\lambda) = \mathcal{O}(N^{-2})$ and thus, recalling Eq.~\eqref{eq:birthtime_clustering}
\begin{equation}\label{eq:clustering_cold_regime}
    \braket{c} = \int_0^1 c(\tau) d \tau = \mathcal{O}(1) \,.
\end{equation}
Note that this asymptotic calculation holds for sparse degree sequence only since $\tilde{\mu}(\tau)$ scales differently for dense networks.

\begin{figure}
    \centering
    \includegraphics[scale=0.6]{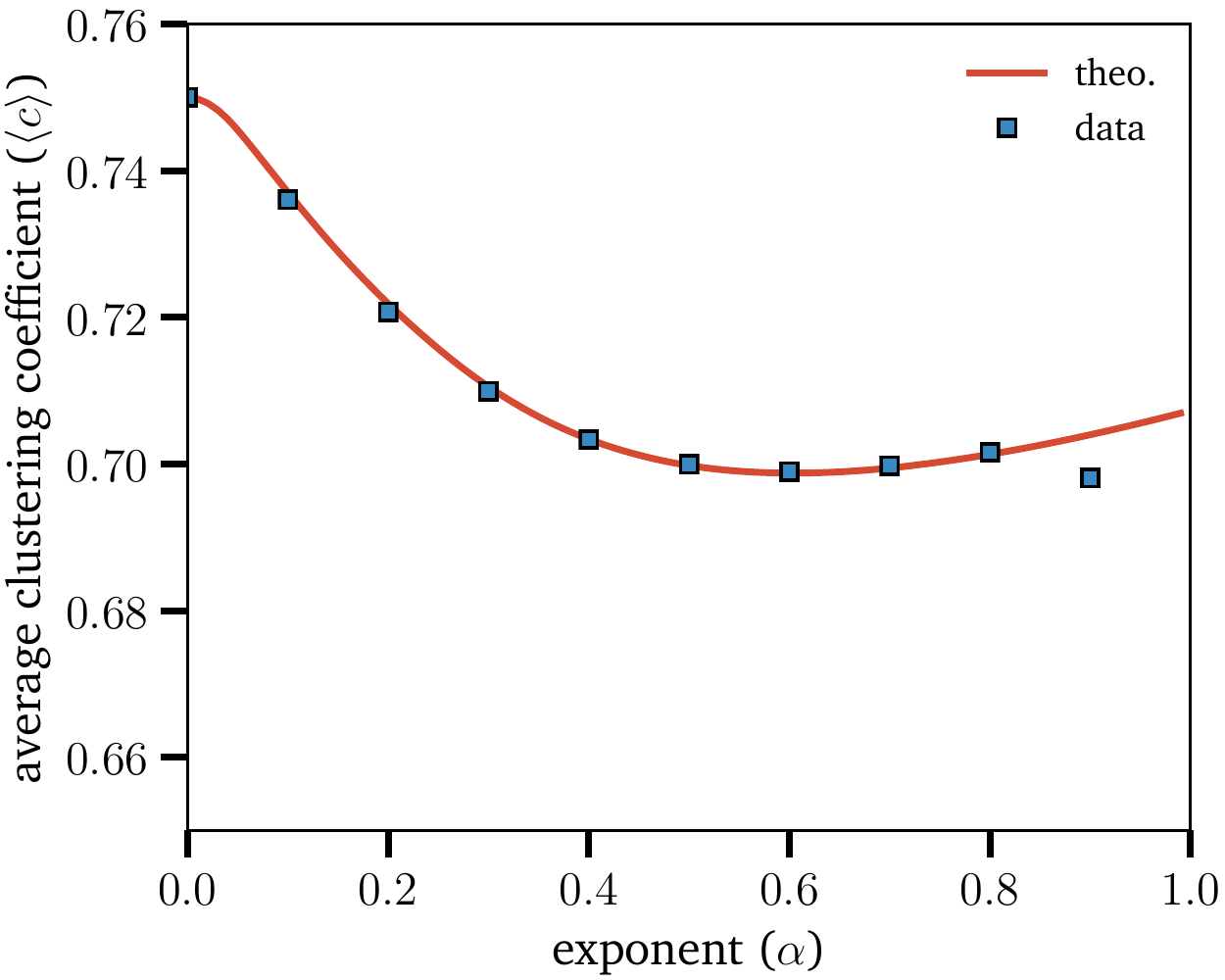}
    \caption{(Color online) Average clustering coefficient $\braket{c}$ as a function of $\alpha$ in the cold regime ($\tilde{\beta}\gg1$). The red line corresponds to the numerical integration of Eq.~\eqref{eq:clustering_cold_regime} and the blue squares, to data from Monte Carlo simulations. We have used $\braket{k} = 50$, $N = 2\times10^4$ and $\tilde{\beta} = 10^6$. The results have been averaged over 96 instances.
    The growing discrepancy at large values of $\alpha$ is strictly due to finite-size effect. }
    \label{fig:c_vs_alpha}
\end{figure}

Let us examine the effect caused by a specific $\kappa(\tau)$. Consider the case of homogeneous degree sequence $\kappa(\tau) = \braket{k}$. Then,
\begin{equation}
    \tilde{\mu}(\tau) \simeq \frac{\braket{k}}{N} \,.
\end{equation}
In that case, we recover the standard RGGs and therefore $\braket{c} = \frac{3}{4}$. For a heterogeneous degree sequence such as Eq.~\eqref{eq:scalefree_degreesequence}, $\braket{c}$ differs from $\frac{3}{4}$ only slightly as seen in Fig.~\ref{fig:c_vs_alpha}.

\subsection{Hot limit $\tilde{\beta}\to0$}

We start by evaluating $\tilde{\mu}(\tau)$ in the hot limit $\tilde{\beta}\to0$,
\begin{equation}
    \tilde{\mu}(\tau) \simeq -\frac{2}{\tilde{\beta}} \ln\br{\frac{1 - n(\tau)}{n(\tau)}} \,.
\end{equation}
Note that, because $\tilde{\mu}(\tau)$ may take negative values, it cannot be interpreted as a connection threshold anymore. While this may seem counterintuitive, it is a necessary condition to preserve the degree sequence. Now, reinjecting this expression in Eq.~\eqref{eq:fermi_dirac_prob} yields,
\begin{equation}\label{eq:hot_connProb}
   p\big[\xi, \psi; \tilde{\mu}(\tau), \tilde{\beta}\big] \simeq n(\tau) \,.
\end{equation}
Interestingly, in the hot limit, the connection probability becomes independent of the position of the nodes. That is to say that the embedding space, and therefore the geometry, does not influence the likelihood of connection between any pair of nodes. Since the connection probability reduces significantly in that limit, $\ell(\tau, \sigma, \lambda)$ is straightforward to obtain
\begin{equation}\label{eq:hot_triangle}
    \ell(\tau, \sigma, \lambda) \simeq n(\tau) n(\sigma) n(\lambda) \,,
\end{equation}
as a product of the three separated connection probabilities. This clearly illustrates the fact that the triangle formations are uncorrelated in this limit.

\begin{figure}
    \centering
    \includegraphics[scale=0.6]{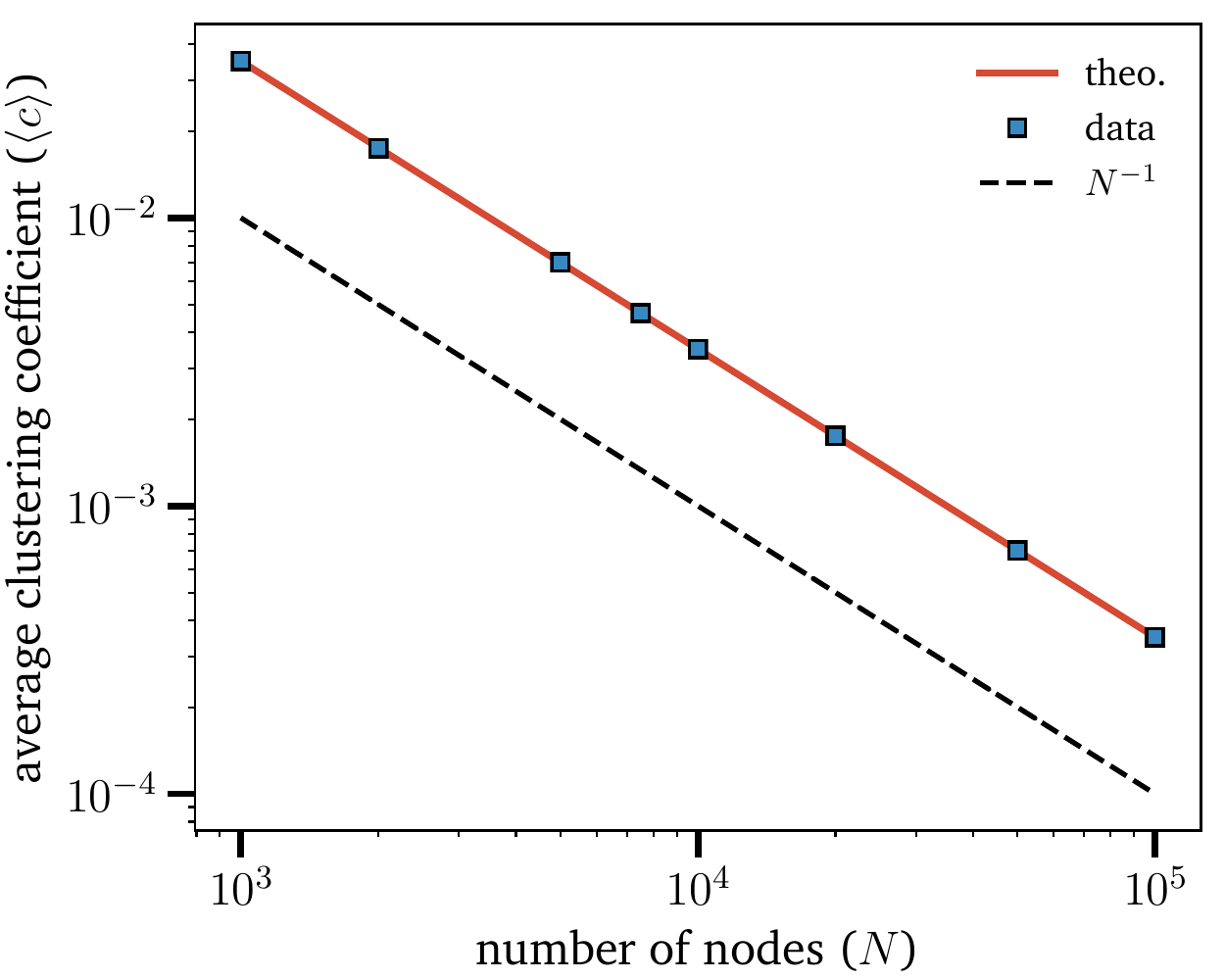}
    \caption{(Color online) Average clustering coefficient $\braket{c}$ as a function of $\alpha$ in the hot regime ($\tilde{\beta}\ll1$). The red line corresponds to the numerical integration of Eq.~\eqref{eq:clustering_hot_regime} and the blue square, to data from Monte Carlo simulations. The scaling of $\braket{c}$ is indicated by the black dashed line. We have used $\alpha=0$, $\braket{k} = 35$ and 
$\tilde{\beta} = 10^{-3}$. The networks have been averaged over 96 instances.}
    \label{fig:c_vs_n}
\end{figure}

By means of Eq.~\eqref{eq:hot_connProb} and \eqref{eq:hot_triangle}, we can determine the scaling of $\braket{c}$ for any $\tilde{\mu}(\tau)$. From Eq.~\eqref{eq:hot_triangle}, we see that $\ell(\tau, \sigma, \lambda) = \mathcal{O}(N^{-3})$ and therefore, recalling Eq.~\eqref{eq:birthtime_clustering}
\begin{equation}\label{eq:clustering_hot_regime}
    \braket{c} = \int_0^1 c(\tau) d \tau = \mathcal{O}(N^{-1}) \,.
\end{equation}
This result is validated by Monte Carlo simulations displayed in Fig.~\ref{fig:c_vs_n} for $\kappa(\tau) = \braket{k}$.

As the average clustering coefficient vanishes in the thermodynamic limit and the connection probability loses its geometry dependence, the networks have also lost their geometric nature. In fact, this network ensemble is a generalization of Erd\"{o}s-Renyi random graphs ensemble $\mathcal{G}(N,p)$ where $p \equiv n(\tau)$ is dependent upon the birth time of nodes. For $\kappa(\tau) = \braket{k}$, we recover the result $\braket{c} = p$ as in the standard $\mathcal{G}(N,p)$.

\subsection{Phase Transition}

\begin{figure*}
    \centering
    \includegraphics[scale=0.6]{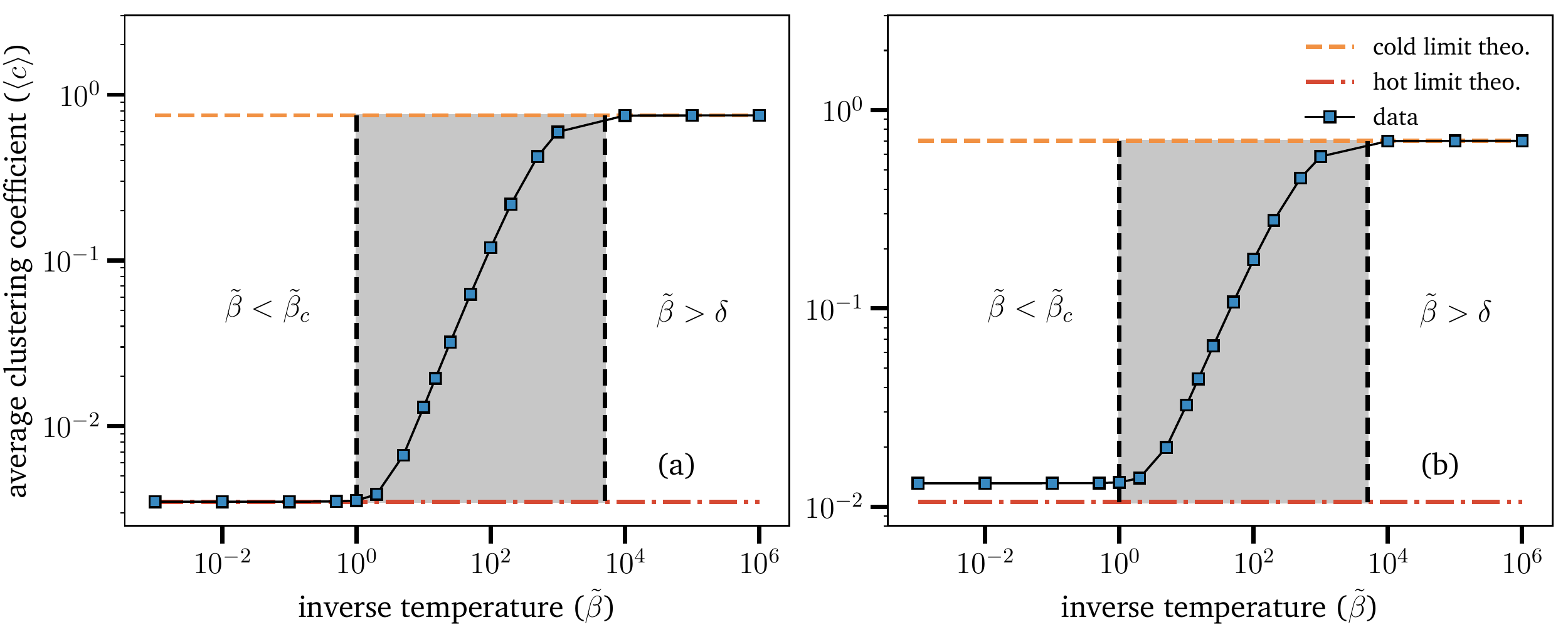}
    \caption{(Color online) Transition between the hot and the cold regime: (a) $\alpha = 0$, $\braket{k} = 35$ and $N= 10^4$, (b) $\alpha = 0.5$, $\braket{k} = 50$ and $N^4$. The orange dashed line (above) corresponds to the cold limit theoretical result while the red dotted line (below) corresponds to the hot limit one. The data from the Monte Carlo simulations is denoted by the squares. We used $N = 10^4$. The networks have been averaged over 20 instances. For $\alpha = 0.5$, noticable finite-size effects are involved in the hot limit, which explains the discrepancy between the theoretical prediction and the simulation data. The hot and cold values of $\braket{c}$
    are respectively $3.5 \times 10^{-3}$ and $0.75$ for (a) and $1.05 \times 10^{-2}$ and $0.70$ for (b) .} 
    \label{fig:c_vs_beta}
\end{figure*}

Varying $\tilde{\beta}$, we observe a phase transition between the random and geometric phases. That is what is shown in Fig.~\ref{fig:c_vs_beta}. Interestingly, the clustering varies between a critical interval $\tilde{\beta}\in[\tilde{\beta}_c, \delta]$, where $\delta \sim \frac{N}{2 \pi \tilde{R}} = \frac{N}{2}$ is the density of nodes on the 2-ball, independent of the degree sequence (see shaded region on Fig.~\ref{fig:c_vs_beta}). That $\braket{c}$ reaches the cold limit when $\tilde{\beta} > \delta$ is due to the saturation of $\braket{c}$.

In the thermodynamic limit, the critical threshold $\tilde{\beta}_c$ is approximately equal to 1, such that $\braket{c} = 0$ for $\tilde{\beta} < \tilde{\beta}_c$, whereas $\braket{c}$ tends asymptotically to the cold limit as $\delta\to\infty$.

%
\section{Conclusion} 
\label{sec:conclusion}

In this paper, we have defined a new type of geometric network growth process in which the newly created links attach homogeneously to the existing nodes. A correspondence between our model and a hidden variable framework has allowed us to determine analytical expressions for the most important structural properties, the degree sequence, the degree-degree correlation and the clustering coefficient. Most importantly, we have shown that the parameter $\tilde{\mu}(\tau)$, the network chemical potential characterizing its geometric evolution, can be used as an adjustable function to reproduce just about any degree sequence. 

Additionally, we have found that the birth time of nodes in the network, characterized by its network history $H$, have a strong influence on the form of the degree-degree correlation. This is perhaps one of the more distinctive features of our model, a result that has not been obtained by existing growth processes. 

Moreover, we have shown that the other parameter $\tilde{\beta}$, the network inverse temperature, allows to interpolate between a random regime, where the connections are not influenced by geometric constraints, and a geometric regime, where these constraints dominate the connection occurrences. The average clustering coefficient $\braket{c}$ varies between two extreme values: the hot limit ($\tilde{\beta}\ll1$) corresponding to the random phase and the cold limit ($\tilde{\beta}\gg1$) corresponding to the geometric phase. Notably, the phase transition between the random and geometric phases with a critical threshold $\tilde{\beta}_c \simeq 1$ is similar to the one found in Ref.~\cite{krioukov2010hyperbolic} in hyperbolic geometry.

Some questions remain open, however. 

First, a problem with certain degree sequences where links are missing when generated by our approach have been identified (see Sec.~\ref{sec:degree_sequence}). We discussed that the allowance of multilinks and self-loops would solve the problem, but this implementation is left for future works. 

Second, it is reasonable to ask whether the ensemble of network generated by our model with given degree sequence and equal weights on all histories yields the network ensemble of the configuration model. This correspondence would be important in a number of ways. On the one hand, it would establish an equivalence with the so-called $\mathbb{S}^1$ model of Ref.~\cite{serrano2008self} and, in turn, with hyperbolic geometry \cite{krioukov2010hyperbolic} and spatial PA \cite{papadopoulos2011popularity,jacob2015spatial}. On the other hand, the ensemble of networks generated by HA would be a generalization of the configuration model network ensemble where an appropriate distribution on the histories can be chosen to reproduce the degree-degree correlations while preserving the degree sequence. 

Third, a further tantalizing question is the possible inference of the (effective) history of a network. Although real networks grow or evolve over time according to their own specific dynamics, our model could nevertheless be used to generate random ensembles of surrogates by reconstructing their \textit{effective} growth history. This could be achieved by inferring the model's parameters through, for instance, the maximization of the likelihood that the model has adequately generated the real network structure. This would yield a network ensemble with similar structural properties (degree sequence, correlations, clustering coefficient). Our preliminary work on this aspect has led to promising results, but much work remains to be done on this front. We expect to report on this venture in the near future. 

%
\section*{Acknowledgements}

The authors are grateful  to J.-G. Young for useful comments and discussions. They acknowledge Calcul Qu\'{e}bec for computing facilities, as well as the financial support of the Natural Sciences and Engineering Research Council of Canada (NSERC), the Fonds de recherche du Qu\'{e}bec--Nature et technologies (FRQNT) and the Canada First Research Excellence Funds (CFREF).
\end{document}